\documentclass[12pt,preprint,a4paper]{aastex}
\usepackage{times}
\usepackage{graphics,epsf}
\usepackage{amsmath}                
\usepackage{amsfonts}               
\usepackage{amssymb}                
\usepackage{epsfig}                 
\usepackage{epstopdf}
\usepackage{rotating}
\usepackage{color}
\usepackage{tabularx}
\usepackage{cancel}

\newcommand{\cm}{{~\rm cm}}
\newcommand{\s}{{~\rm s}}
\newcommand{\ms}{{~\rm ms}}
\newcommand{\km}{{~\rm km}}
\newcommand{\g}{{~\rm g}}
\newcommand{\K}{{~\rm K}}
\newcommand{\erg}{{~\rm erg}}
\newcommand{\yr}{{~\rm yr}}

\def \apj{ApJ}
\def \aap{A\&A}

\def \mnras{MNRAS}
\def \apjl{ApJ Lett.}

\def \nat{Nature}

\def \na{New A}

\begin{document}

 \title{Launching jets from accretion belts}


 \author{Ron Schreier and Noam Soker\altaffilmark{1}}
 \altaffiltext{1}{Department of Physics, Technion -- Israel Institute of Technology, Haifa 32000 Israel; ronsr@physics.technion.ac.il, soker@physics.technion.ac.il.}

\begin{abstract}
We propose that sub-Keplerian accretion belts around stars might launch jets. The sub-Keplerian inflow does not form a rotationally supported accretion disk, but it rather reaches the accreting object from a wide solid angle.  The basic ingredients of the flow are a turbulent region where the accretion belt interacts with the accreting object via a shear layer, and two avoidance regions on the poles where the accretion rate is very low.
A dynamo that is developed in the shear layer amplifies magnetic fields to high values. It is likely that the amplified magnetic fields form polar outflows from the avoidance regions. Our speculative belt-launched jets model has implications to a rich variety of astrophysical objects, from the removal of common envelopes to the explosion of core collapse supernovae by jittering jets.
\end{abstract}


\section{INTRODUCTION}
\label{sec:introduction}

Jets are known to be launched by accretion disks around compact objects (e.g., \citealt{Livio2011}), such as super-massive black holes in active galactic nuclei, X-ray binaries, young stellar objects (YSOs), and in planetary nebulae (PNe; e.g., \citealt{SahaiNyman2000}).
{{{{ In many of these objects the jets are observed and resolved. We here proposed a mechanism to launch jets by objects that are embedded inside optically thick gas, and in most cases cannot be observed. In such cases the gas originates from regions close to the compact object, and in some cases does not have large enough specific angular momentum to form an accretion disk. Instead, the accreted gas forms a sub-Keplerian inflow with density increasing toward the equator of the compact object. We term this flow an accretion belt. We are particularly motivated to shed light on the possibility that jets are launched by accretion belts around main sequence (MS) stars that preform a spiraling-in motion inside the large envelope of a giant star; i.e., a common envelope evolution (CEE).
In such a situation the mechanism of jet launching should not involve an extended accretion disk. As such, we also relax the assumption that large scale magnetic fields exist in the accreted gas.
}}}}

Many jet launching models are based on the operation of large scale
magnetic fields, i.e., those with coherence scale larger than the
radius of the disk at the considered location (e.g.,
\citealt{ZanniFerreira2013} and \citealt{Narayanetal2014} and
references therein). Some models do not rely on the magnetic
fields of the accreting compact object, but rather  assume an
outflow from an extended disk region with large-scale magnetic
fields (e.g., \citealt{KoniglPudritz2000, Shuetal2000,
Ferreira2002, Krasnopolskyetal2003, FerreiraCasse2004, Murphyetal2010, Sheikhnezamietal2012, Tzeferacosetal2013, SheikhnezamiFendt2015}).

{{{{ We note the following properties that support our approach of not basing our mechanism on large scale magnetic fields. }}}} Firstly, the strong dynamo that operates in accretion disks is likely to modify the structure  of the large scale magnetic fields, (e.g. \citealt{Hujeiratetal2003}). Secondly, \cite{Ferreira2013} concluded that steady state jets' launching models cannot spin down protostars to the observed values of roughly $10 \%$ of their break-up speed. In contrast, unsteady models are better candidates for the removal of energy and angular momentum from the disk.
{{{{ Our proposed mechanism for jet launching in sub-Keplerian accretion flows is based on unsteady behavior of the magnetic fields. Our proposed mechanism incorporates reconnection of magnetic field lines that takes place in many sporadic events in the accretion belt, much as magnetic activity occurs on the solar surface. These events, we propose in the study, eject mass along the polar directions and lead to the formation of jets. }}}}

We aim at two cases in particular: (1) The case where a MS star is spiraling-in inside the envelope of a giant star. The MS star can accrete mass from the envelope at a rate of the order of the Eddington limit. The specific angular momentum of the accreted gas, $j_{\rm acc}$, might be in many cases below that required to form a Keplerian accretion disk around the MS star, $j_{\rm Kep}$, but still be non-negligible $j_{\rm acc} \approx 0.1-1 j_{\rm Kep}$ \citep{Soker2004}.
(2) An accretion by the newly born NS in core collapse supernovae (CCSNe). In some CCSNe the accreted gas might possess a rapidly varying non-negligible value of the specific
angular momentum, but not sufficient to form a Keplerian accretion
disk \citep{GilkisSoker2015, GilkisSoker2016, Papishetal2016}.
Both scenarios are most difficult for observational validation.

Incorporating ingredients from accretion disks formed at high accretion rates we build the belt-launched jets model. The basic ingredients of the model are described in section \ref{sec:ingredients}, and the dynamo is discussed in section \ref{sec:dynamo}. The implications for some specific astrophysical objects are discussed in section \ref{sec:summary} that also contains our summery.

\section{BASIC INGREDIENTS}
\label{sec:ingredients}

\begin{figure}
    \centering
  \includegraphics[width=0.7\textwidth]{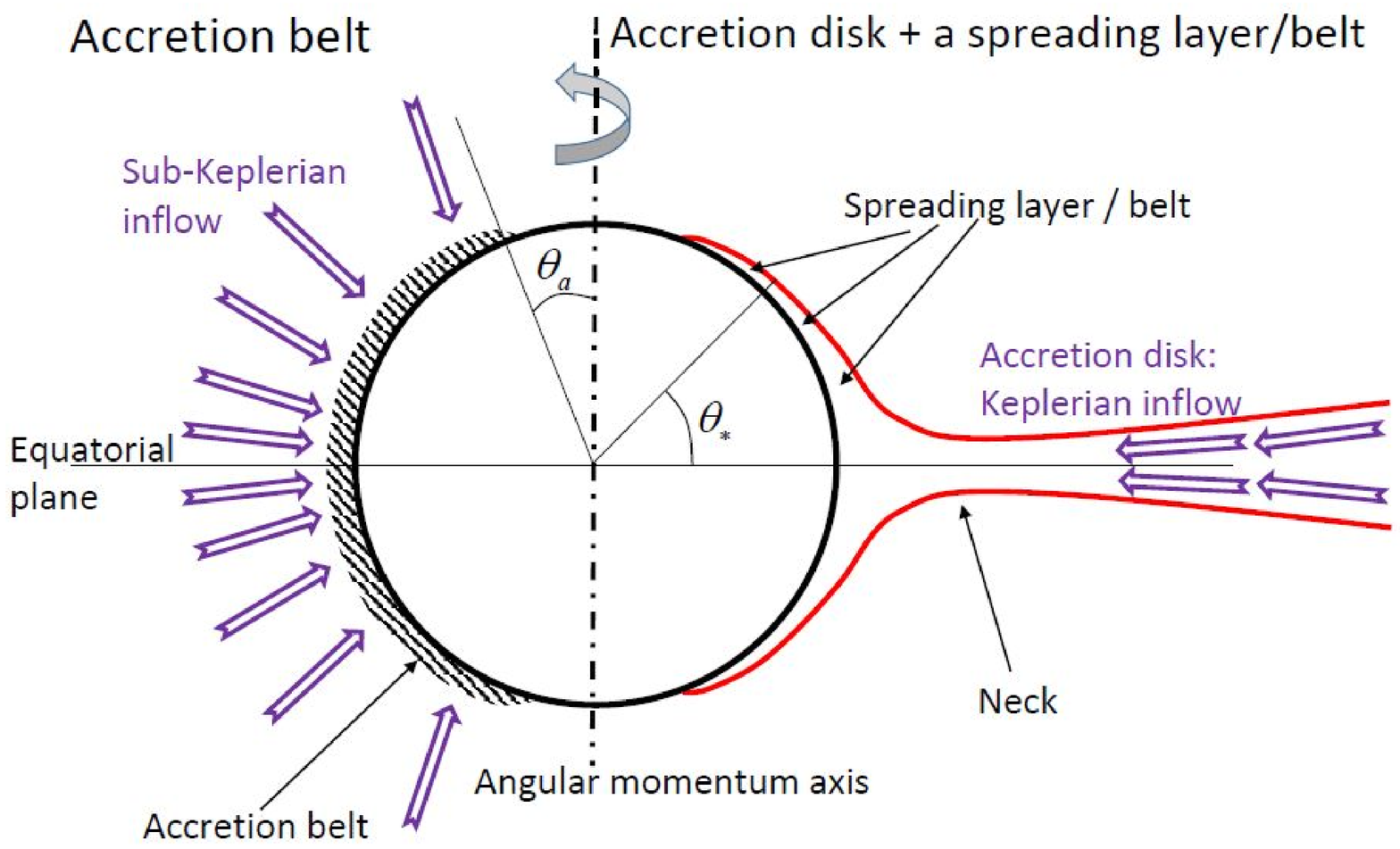}
\caption{The flow structure studied here is presented schematically in the left half of the figure. It has some similarities with the spreading-layer that was proposed by \citep{InogamovSunyaev1999} that is presented on the right half. The meridional extent of the hot part of the spreading
layer $\theta_\star$ is given by equation (\ref{eq:angle}). The angle of avoidance region $\theta_a$ is given by equation (\ref{eq:limitangle}).   }
    \label{fig:flow1}
\end{figure}

A geometrically-thin Keplerian accretion disk is connected to the
accreting body, if it is not a black hole and if the accreting body magnetic field
is not too strong, via a thin high-shear layer termed the boundary
layer (e.g., \citealt{Regev1983, RegevBertout1995}). The
matter in the boundary layer expands to higher latitudes along the
surface of the accreting body. \cite{InogamovSunyaev1999} built a
model for cases of high accretion rates onto neutron stars (NS), and
proposed that the meridional spreading brings the accreted mass
close to the poles. In that model the luminosity from the
spreading-layer, or belt, reaches a maximum away from the
equatorial plane. The accreted gas spins-down to the rotation
velocity of the accreting body through a turbulent friction that
exists in the spreading-layer \citep{InogamovSunyaev1999,
InogamovSunyaev2010}. This type of flow is depicted in the right
side of Fig. \ref{fig:flow1}. The left half-side of Fig.
\ref{fig:flow1} schematically presents the flow structure proposed
in the present study.

{{{{ Two comments are in place here with regards to Fig. \ref{fig:flow1}. ($i$) In this study we consider cases where a Keplerian disk does not form, at least not initially. It is true that when the thermal and/or magnetic pressures are considered the angular velocity of the gas in the disk is slightly sub-Keplerian. For example, simulations for jets-launching disks have slightly sub-Keplerian disks (e.g., \citealt{CasseKeppens2002, Zannietal2007}). However, we consider cases where the magnetic and thermal pressure are not sufficient to allow an accretion disk at early times. Later, after the initial magnetic field has been amplified and viscosity transferred angular momentum outward, an accretion disk might be formed. If an accretion disk does form, then the launching of jets becomes even more likely. We here take the pessimistic case that an accretion disk does not form.
($ii$) Because of the sub-Keplerian angular momentum, away from the accreting star the flow is not influenced much by the rotation. For example, for a MS star orbiting inside a giant star the accretion might be that of Bondi-Hoyle-Lyttleton accretion. For a NS accreting from a collapsing core, the flow can be spherical on large scale. However, when the angular velocity is sub-Keplerian but non-negligible, very close to the surface of the star the gas is concentrated toward the equatorial plane. In the polar regions the density of the inflowing gas is very low.  }}}}

It is not possible to directly apply the results of
\cite{InogamovSunyaev1999} to the cases we study here (listed in
section \ref{sec:introduction}). We study MS stars
accreting at very high rates, a process that is different than
accretion onto a NS. For example, radiation pressure might not be
as important as in accretion onto a NS. In the flow structure onto
a newly born NS in CCSNe, cooling is mainly via neutrinos, and not
by photon emission. The flow structures we discuss here are more
complicated, and 3D numerical studies will be required to simulate
such flows. The study of \cite{InogamovSunyaev1999} non the less
suggests that an energetic wide accretion belt can be formed on the
surface of the accreting body, and that turbulence is expected in
such a belt.

Neglecting magnetic fields, \cite{InogamovSunyaev1999} find that
the meridional extent of the hot part of the spreading layer (the
belt) $\theta_\star$, is approximately given by
 \begin{equation}
 \sin \theta _\star \approx \frac {\dot M_{\rm acc} v^2_K}
 {2  L_{\rm edd} }  ,
  \label{eq:angle}
 \end{equation}
where $L_{\rm edd}$ is the Eddington luminosity limit, $\dot
M_{\rm acc}$ is the accretion rate, and $v_K$ is the Keplerian
velocity on the surface of the accreting body. One can find that
for accreting rates larger than about $10^{18} \g \s^{-1} \simeq
10^{-8} M_\odot \yr^{-1}$ onto a NS with a mass of $ M_{\rm NS}
\approx 1.4 M_\sun$ and a radius of $R=12 \km$, the spreading
layer extends to the polar regions. We here also apply equation
(\ref{eq:angle}) to accretion onto a MS star. We find
that for the spreading layer to reach the polar region of an
accreting solar-like star, the accretion rate should be larger
than approximately $10^{-3} M_\odot \yr^{-1}$.  Such accretion rates can be achieved during the common envelope evolution.

We note that equation (\ref{eq:angle}) was developed for a belt formed by an
accretion disk. We here investigate cases where no
geometrically-thin accretion disk is formed, but rather the
accreted gas streams from a wide angle, with an avoidance region
near the poles.
{{{{ As well, equation (\ref{eq:angle}) does not take into account magnetic activity. The magnetic activity leads to mass ejection from the polar regions, and maintain that region open to an outflow (see below). }}}}

In the sub-Keplerian inflow scenario the accreted gas has an
average specific angular momentum of $j_{\rm acc} < j_{\rm Kep} $,
where  $j_{\rm Kep} = (G M R)^{1/2}$ is the Keplerian specific angular momentum
on the surface of the accreting object of mass $M$ and radius $R$.
A parcel of gas with such an angular momentum cannot be accreted
to the surface of the star within an angle of $\theta_a$ from the
poles, given by \citep{Papishetal2016}
\begin{equation}
\theta_a = \sin^{-1} \sqrt{\frac{j_{\rm acc}}{j_\mathrm{Kep}}}.
\label{eq:limitangle}
\end{equation}
This is termed the avoidance angle.

In section  \ref{sec:dynamo} we speculate that the extended belt
formed in the sub-Keplerian accretion flow can launch jets. We
suggest there that the strong radial shear due to the powerful
differential rotation together with the turbulence, amplify
magnetic fields through the dynamo action. MHD effects, such as reconnection of magnetic field lines and dragging gas in the polar directions via magnetic stress of filed lines that are anchored to the accreted gas, then launch a polar-collimated outflow. The point to make here is that in the
case of a belt formed by an accretion disk it is expected that
more mass will be launched by the accretion disk than by the belt.
In the sub-Keplerian wide inflow case, most, or even all, of the
outflowing gas is launched by the belt.

{{{{ The amplification of magnetic fields affects the opening angle as given in equation (\ref{eq:limitangle}). ($i$) Reconnection of the magnetic field lines eject gas through the two opposite polar avoidance regions. This activity, that leads to the formation of an outflow and high magnetic pressure region, might increase the opening angle in the inflowing gas. ($ii$) Winding of the magnetic field lines frozen to the polar outflow channels rotation energy to outflow kinetic energy. The  magnetic tension of the magnetic field lines might further increase the opening angle. Over all, the opening angle increases with the specific angular momentum of the accreted gas, and might be further increased by magnetic activity. }}}}

The basic ingredients of the belt-launched jets scenario are presented in Table 1 alongside with those of jets launched by accretion disks. Our proposed launching mechanism does not require large scale magnetic fields and does not require a thin Keplerian disk. The most important ingredient is the operation of a dynamo, the belt dynamo.
\begin{table}[ht]
 \centering \caption{Jet launching cases}
 \begin{tabular}{c c c} 
 \hline
 Physical parameter      &      Accretion disk          &  \textbf{Belt}  \\
 [0.5ex]
 \hline \hline
 Accretion rate ($\dot M_{\rm acc}$)&up to $\approx \dot M_{\rm Edd}$& $\ga \dot M_{\rm Edd}$  \\
 \hline
Accreted specific angular& $ \gg j_{\rm Kep}$           &   $ < j_{\rm Kep}$  \\
momentum ($j_{\rm acc}$) &                              &      \\
 \hline
Angular velocity         & $\approx \Omega_{\rm Kep}$   & $< \Omega_{\rm Kep}$   \\
at launching ($\Omega_ {\rm L}$)  &                     &      \\
 \hline
 Launching area ($D_{\rm L}$)  &  $\gg R$               &    $\simeq R$ \\
 \hline
 Magnetic fields        & weak to strong   & $ \vec B$ is amplified       \\
                        &                  & locally via dynamo       \\
 \hline
 Jet's energy          & Gravitational of     &  Gravitational of \\
  source               & accreted gas         &  accreted gas     \\
  \hline
  \end{tabular}
\label{table:nonlin} 
\newline
 The different symbols have the following meaning: $R$ is the
 radius of the accreting body; $\dot M_{\rm Edd}$ is the Eddington
 accretion limit; $j_{\rm Kep}$ and $\Omega_{\rm Kep}$ are the Keplerian specific angular momentum and angular velocity very close to the accreting body surface, respectively.
\end{table}

\section{THE BELT DYNAMO}
 \label{sec:dynamo}

There are several approaches to estimate the amplification of magnetic fields in sheared layers in stellar interiors, both in convective and non-convective regions. In the present preliminary study we take one calculation referring a non-convective region. We then argue that in a convective regions of the accretion belt the amplification will be more efficient even.

\cite{Spruit2002} gave the analytic ground for quantifying the magnetic fields created in non-convective layers in a differentially rotating star. The differential rotation stretches poloidal field lines into toroidal fields. Magnetic instabilities in the amplified toroidal magnetic fields replace the role of convection in creating more poloidal magnetic field from the toroidal field. This process, that is relatively slow in stellar interior, is expected to be more efficient in the more rapidly spinning belt. The turbulent regions of the belt are even more efficient in amplifying the magnetic field. We turn to show that even in the non-turbulent regions of the belt the dynamo can be effective.

Due to the strong differential rotation (shear) occurring in the belt the radial component
of the field, $B_r$, is twisted into the azimuthal
direction, $B_\phi$, so that after few rotations
it becomes the dominant component.
Its strength increases linearly with time, until it becomes unstable.
The amplitude of the dynamo-generated field in the non-turbulent regions is given by \citep{Spruit2002}
\begin{equation}
B_\phi  =
 r\,(4\pi\rho)^{1/2}\,\Omega_{\rm belt}\, q^{1/2}
 \left(\frac{\Omega_{\rm belt}}{N}\right)^{1/8}\left(\frac{\kappa}{r^2N}\right)^{1/8}
\label{eq:Bphi}
\end{equation}
\begin{equation}
B_r =  B_\phi\left(\frac{\Omega_{\rm belt}}{N}\right)^{1/4} \left(\frac{\kappa}{r^2N}\right)^{1/4},
\label{eq:Br}
\end{equation}
where the density in the belt is estimated from mass conservation of the inflowing gas
\begin{equation}
\rho = \dot{M} \left(4\pi r^2\,v_{\rm in} \right)^{-1} ,
\label{eq:bldensity}
\end{equation}
and $v_{\rm in}$ is the radial inflow speed into the belt, which we take as the free-fall velocity.
The temperature of the accreted plasma when it is stopped on the accreting object is calculated from $3/2 kT\, = 1/2\,m_p v_{\rm in}^2$.
Here $\Omega_{\rm belt}$ is the belt angular velocity, and $N$ is the buoyancy (Brunt-Vaisala) frequency that is slightly less than the Keplerian frequency. We can crudely take
$N \approx \Omega_{\rm belt} $.   The thermal diffusivity is
$\kappa=16\sigma T^3/(3\kappa_R\rho^2 c_p)$, where $c_p$ is the heat capacity per unit mass, $\kappa_R$ is the opacity, and $q=(r\partial_r\Omega_{\rm belt})/\Omega_{\rm belt} \simeq 1- 3$ is the dimensionless differential rotation rate.

As the cooling in accreting NS in CCSNe is due to neutrinos, we apply (\ref{eq:Bphi}) and (\ref{eq:Br}) to accreting MS stars.  We take an accretion rate of $\dot M_{\rm acc} = 10^{-3} M_\odot \yr^{-1}$ onto a solar like star, from which we find the density of the accretion inflow near the surface, and the temperature of the gas in the belt to be $T=10^6 \K$. We find that $\kappa \approx 10^{22} \cm^2 \s^{-1}$, and so
$[\kappa/(r^2N) ]^{1/8} \approx 1$.
We can derive from (\ref{eq:Bphi}) the ratio of the magnetic field energy density to that of the kinetic and thermal gas density, for an accreting MS star at a rate of $\dot M_{\rm acc} \approx 10^{-2} -10^{-4} M_\odot \yr^{-1}$. For the gas energy density we take $\rho v^2_{\rm esc}/2$, where $v_{\rm esc}$ is the escape velocity from the star
\begin{eqnarray}
\frac {B^2/4 \pi }{e_{\rm gas}} \approx 0.2   q
\left( \frac {\Omega_{\rm belt}/\Omega_{\rm Kep}}{0.3} \right)^2
\left(\frac{\Omega_{\rm belt}}{N}\right)^{1/4}\left(\frac{\kappa}{r^2N}\right)^{1/4} .
\label{eq:Bscaled}
\end{eqnarray}
As stated, this is for the non-turbulence part of the accretion belt. As turbulence is expected in the strongly sheared belt, the amplification can be more efficient event. We conclude that a fraction of $\approx {\rm few} \times 0.1$ of the accreted energy is channelled to magnetic energy. In our proposed scenario the magnetic field lines are further winded and stretched by the rotating belt as they are dragged by the outflowing gas. The magnetic fields reconnect and launch jets.

In claiming that an efficient dynamo can be operate in the accretion belt we are encouraged by the recent results of \cite{Mostaetal2015} who conducted very high resolution simulations of CCSNe with pre-collpase rapidly rotating cores. They show that rapidly rotating material around the newly born NS can amplify tremendously an initial magnetic field, and leads to jets lunching.
In their simulations the turbulent kinetic energy in the accreted gas is converted into electromagnetic energy. The timescale for an e-fold increase in the magnetic field in the accretion disk, $\tau_e \approx 0.5 \ms$, is about half an orbital period in the relevant part of the disk. Interestingly, only very-high resolution simulations were able to demonstrate the tremendous magnetic field amplification. These results are very supportive for the jittering-jets model for the explosion of all CCSNe with explosion energies of $\ga 2 \times 10^{50} \erg$.

Intermittent accretion belts in CCSNe are expected to last for a time of $\tau_b \approx 0.01-0.1 \s$, {{{{  as estimated either from the fluctuation in the convective shells of the pre-collapse core \citep{GilkisSoker2016}, or from fluctuations in the post-collapse region around the NS formed by the standing accretion shock instability (SASI; \citealt{Papishetal2016}).   }}}}
This time scale is about equal to tens of Keplerian orbits at $\sim 20 \km $ from the NS. A belt with a specific angular momentum of $\eta \equiv j_{\rm acc}/j_{\rm Kep}$ has an orbital period on the equator of the newly born NS of $\approx 26 (\eta/0.05)^{-1} (r/20 \km)^{3/2} \ms$. Closer to the poles the period will be shorter. {{{{ Over all, the accretion belt can last for a time scale that is several times and up to tens times a half orbital period. According to the results of \cite{Mostaetal2015} this is a sufficiently long time to amplify the magnetic fields. }}}}
This shows that the sub-Keplerian disk might have sufficient time to substantially amplify the magnetic fields. We conjecture that the strong magnetic fields with the preferred axis of rotation and the avoidance angle (eq. \ref{eq:limitangle}), lead to the launching of two opposite jets in the polar directions, where the ram pressure of the accreted gas is very low.

\section{DISCUSSION AND SUMMARY}
 \label{sec:summary}

We conducted a preliminary study that led us to argue that sub-Keplerian accretion flows onto compact objects can launch jets. The sub-Keplerian accreted gas forms an accretion belt rather than an accretion disk (left half side of Fig. \ref{fig:flow1}). Within the avoidance angle $\theta_a$ from the polar directions (eq. \ref{eq:limitangle}) the accretion rate is very low. The dynamo amplification of the magnetic fields within the belt (section \ref{sec:dynamo}) can lead to very strong magnetic fields, as can be seen from equation (\ref{eq:Bscaled}) that gives the ratio of the amplified magnetic field energy density to that in the accreted gas.

We speculate that reconnection of the magnetic field lines can lead to an outflow through the two opposite polar avoidance regions. Winding of the magnetic field lines frozen to the polar outflow  can further channel rotation energy to outflow kinetic energy.
The main differences and common properties of the proposed belt-launched jets scenario from those of the common disk-launched jets are presented in Table 1.

Our conclusion is that accretion belts that are formed by sub-Keplerian accretion flows might lunch jets. If true, these results might have far reaching implications for the removal of common envelopes by jets. Numerical studies point at a limited efficiency of the common envelope removal by the gravitational energy released during the spiraling-in process (e.g.,
\citealt{DeMarco2011, Passyetal2012, RickerTaam2012, Ohlmannetal2015}).
It has been suggested that jets can assist it removing the common envelope (e.g. \citealt{Soker2014}). However, the specific angular momentum of the gas accreted by a MS star spiraling-in inside a giant envelope is sub-Keplerian $j_{\rm acc} \approx 0.1-1 j_{\rm Kep}$  \citep{Soker2004}. A belt is expected to be formed around the MS star during the
common envelope phase. Jets that might be launched by the belt, as argued in the present study, can assist in removing the common envelope.

As well, our results might have implications to CCSNe explosion scenarios. The neutrino-delayed mechanism has severe problems in accounting for explosions with kinetic energy of more than about
$2-5 \times 10^{50} \erg$ \citep{Papishetal2015}. An intermittent accretion belt is expected to be formed around the newly born NS during the first several seconds of the explosion
\citep{GilkisSoker2015, GilkisSoker2016, Papishetal2015}. If the present results hold to that flow structure, then the jets can explode the star, up to explosion energies of above $10^{52} \erg$
\citep{Gilkisetal2016}.  We note that because of the stochastic nature of the angular momentum of the accreted mass in many CCSNe, in the jittering jets model the spin of the NS might be inclined
to the momentarily angular momentum of the belt. The effect of this will have to be studied in future numerical simulations.

In both cases, of common envelope removal and CCSNe, the origin region of the jets is completely  obscured. In cases where the origin of jets is seen, such as in young stellar objects, active galactic nuclei, and some binary systems, the accretion rate is low and an accretion disk is required to be formed. We here argued that in cases where accretion rate is very high, hence the entire region is heavily obscured, even accretion belt formed by sub-Keplerian inflow can launch jets (see also \citealt{Shiberetal2016}).

We can summarize our study by stating that the possibility for accretion belts to launch collimated outflows, or jets, opens a rich variety of processes that can account for some puzzles in astrophysics, such as the explosion of massive stars and in some cases in the removal of the common envelope.

{{{{ We thank an anonymous referee for valuable comments. }}} }
This research was supported by the Asher Fund for Space Research at the Technion, and the E. and J. Bishop Research Fund at the Technion.

{}

\end{document}